\documentclass[twocolumn,english]{revtex4-1}
\usepackage[latin9]{inputenc}
\usepackage{verbatim}
\usepackage{units}
\usepackage{amsmath}
\usepackage{amssymb}
\usepackage{graphicx}
\usepackage{babel}
\begin{document}
\global\long\def\ket#1{\left|#1\right\rangle }

\global\long\def\bra#1{\left\langle #1\right|}

\global\long\def\braket#1#2{\left\langle #1\left|#2\right.\right\rangle }

\global\long\def\ketbra#1#2{\left|#1\right\rangle \left\langle #2\right|}

\global\long\def\braOket#1#2#3{\left\langle #1\left|#2\right|#3\right\rangle }

\global\long\def\mc#1{\mathcal{#1}}

\global\long\def\nrm#1{\left\Vert #1\right\Vert }

\global\long\def\unit{1\!\!1}

\title{Quantum heat machines equivalence and work extraction beyond Markovianity,
and strong coupling via heat exchangers }

\author{Raam Uzdin}

\email{raam@mail.huji.ac.il}

\author{Amikam Levy}

\author{Ronnie Kosloff}

\affiliation{Fritz Haber Research Center for Molecular Dynamics, Hebrew University
of Jerusalem, Jerusalem 9190401, Israel}
\begin{abstract}
Various engine types are thermodynamically equivalent in the quantum
limit of small ``engine action\textquotedblright . Our previous derivation
of the equivalence is restricted to Markovian heat baths and to implicit
classical work repository (e.g., laser light in the semi-classical
approximation). In this paper all the components, baths, batteries,
and engine, are explicitly taken into account. To neatly treat non-Markovian
dynamics we use mediating particles that function as a heat exchanger.
We find that on top of the previously observed equivalence there is
a higher degree of equivalence that cannot be achieved in the Markovian
regime. Next we focus on the energy transfer to the work repository.
A condition for zero entropy increase in the work repository is given.
Moreover, it is shown that in the strong coupling regime it is possible
to charge a battery with energy while reducing its entropy at the
same time. 
\end{abstract}
\maketitle
All heat engines, classical and quantum, extract work from heat flows
between at least two heat baths. When the working fluid is very small
and quantum , e.g., just a single particle, the dynamics of the engine
can be very different from that of classical engines \cite{EquivPRX,MitchisonHuber2015CoherenceAssitedCooling}.
Nonetheless, some classical thermodynamic restrictions are still valid.
For example, quantum heat engines are limited by the Carnot efficiency
even when the dynamics is quantum. Today, it is fairly well understood
why the Clausius inequality originally conceived for steam engines
still holds for small quantum heat machinea. 

The field of quantum thermodynamics is intensively studied in recent
years. The main research directions are the study of quantum heat
machines, thermodynamic resources theory, and the emergence of thermal
states. See the recent reviews \cite{goold2015review,SaiJanetReview,millen2015review}
and references therein for more information on these research directions. 

The study of quantum heat machines dates back to \cite{scovil59}
where it was shown that the lasing condition for a system pumped by
two heat baths corresponds to the transition from refrigerator to
an engine. See \cite{k102,k122} for a more detailed analysis of such
systems. Since then, various types of heat machines have been studied:
reciprocating, continuous, autonomous and non autonomous, four-stroke
machines, two-stroke machines, Otto engines and Carnot engines in
the quantum regime. See \cite{alicki79,k24,k152,k221,levy14,rahav12,allmahler10,linden10,mahler07b,skrzypczyk2014work,gelbwaser13,kolar13,alicki2014quantum,Nori2007QHE,lutz14,BinderOperational,Correa2014EnhancedSciRep,dorner2013extracting,palao13,dorner2012emergent,DelCampo2014moreBang,malabarba2014clock,Gelbwaser2015Rev,WhitneyThermElec,AllahverdyanOptDualStroke,mari2012,AlgoCoolQuantGas2011Nature,AlgoCool2005expNature,boykin2002algorithmic,mahler07,segal06,schulman2005physical,tal02,PopescuSallMaxEff,SeifertCohFeedbackEng,EspositoFastDriven,MartiWorkCorr}
for a rather partial list on heat machine studies in recent years.

Experimentally, a single ion heat engine \cite{rossnagelIonEngExp}
and an NMR refrigerator \cite{AlgoCool2005expNature} have already
been built. Suggestions for realizations in several other quantum
systems include quantum dots \cite{QuantDotEngine,BergenfeldtQDotsMicrowave},
superconducting devices \cite{PekolaSCengine,HuberEntangGenHeat,campisi2014FT_SolidStateExp,pekola2015towards},
cold bosons \cite{FialkoColdBosonsEng}, and optomechanical systems
\cite{mari2012,ZhangOptoMechEng,OptNanomechEng}.

The second law was found to be valid for heat machines \cite{alicki79}
in the weak system-bath coupling, where the Markovian dynamics is
described by the Lindblad equation. In fact, the second law is consistent
with quantum mechanics regardless of Markovianity as long as proper
thermal baths are used \cite{sagawa2012second}. One of the main the
challenges in this field is to find 'quantum signatures' \cite{EquivPRX}
in the operation of heat machines. More accurately, to find quantum
signatures in \textit{thermodynamic} quantities such as work, heat,
and entropy production. Clearly, the engine itself is quantum and
as such it may involve quantum features such as coherences and entanglement.
The question is whether by measuring only thermodynamics quantities
such as average heat or work, it is possible to distinguish between
a quantum engine and a classical one. %
\begin{comment}
For example think of two friends one is an experimentalist the other
is a theorist. The experimentalist tells his friend he built an engine
and ask him to tell if it is quantum or not by looking at the thermodynamic
inputs and outputs of the machine. The theorist can ask to reduce
or increase any parameters of the engine e.g. the cycle time, the
system-bath coupling time the temperatures and get new measurements
with the new parameters. Recently it was shown that {[}{]} the theorist
can win this games
\end{comment}
{} 

As it turns out, there are thermodynamic effects that are purely quantum,
the most relevant to this work is the equivalence of heat machine
types \cite{EquivPRX}. Other quantum thermodynamic effects include
extraction of work from coherences \cite{Anders2015MeasurementWork}
and oscillation in cooling \cite{MitchisonHuber2015CoherenceAssitedCooling}.
In resources theory it seems that quantum coherences in the energy
basis also play an important role, and impose restrictions on the
possible single shot dynamics \cite{LostaglioRudolphCohConstraint,OppenheimCohEvoLim}.

The traditional models and analysis of quantum heat machines resemble
that of laser physics in the semi classical approximation. The driving
field is often modeled by a classical electromagnetic field. This
field generates a time-dependent Hamiltonian so it is possible to
extract pure work from the system. When the classical field is replaced
by a work repository (battery) with quantum description the dynamics
becomes more complicated \cite{KamilCohWork}. For example, for an
harmonic oscillator battery, the initial state of the battery has
to be fairly delocalized in energy to avoid entropy generation in
the baths \cite{aaberg2014catalytic,malabarba2014clock}. This is
problematic since an oscillator always has a ground state. See \cite{KamilCohWork}
for a detailed account of this mechanism. In this work we shall use
multiple batteries to extract work by interacting with the engine
via energy conserving unitary evolution (ECU). Interestingly, machines
without classical fields have been previously studied \cite{linden10,k272,malabarba2014clock}.
However, the research goals in these studies are entirely different
from those of the present study.

Another assumption that is almost always used in the analysis of heat
machines, is that of weak coupling to the bath. Weak coupling, initial
product state assumption, and other approximations lead to the Lindblad
equation for the description of the thermalization process. The Lindblad
equation is widely used in open quantum systems and they describe
very well the dynamics in many scenarios. In other scenarios, such
as strong system-bath interactions, or for very short evolution times,
the Lindblad equation fails \cite{breuer}. In the scheme presented
in this paper we include heat exchangers. Their role is to enable
non-Markovian engine dynamics while still using Markovian baths. 

One of the goals of this paper is to show that heat machine equivalence
goes beyond the classical field approximation and also for very short
times where the Markovian approximation does not hold. 

In a three-body engine all these interactions occur simultaneously.
Consequently, three-body machines are not suited for describing reciprocating
machines and their effects.

A 'stroke' of a quantum machine is defined in the following way \cite{EquivPRX}.
It is an operation that takes place in a certain time segment. In
addition, a stroke does not commute with the operations (strokes)
that take place before or after it. This non commutativity is essential
for thermodynamic machines. Without it the system will reach a state
that is compatible with all baths and batteries, and no energy flows
will take place. Different machine types differ in the order of the
non commuting operations. In a two-stroke machine, the first stroke
generates simultaneous thermalization of two different parts of the
machine (manifolds) to different temperatures. In the next stroke,
a time dependent Hamiltonian couples the two manifolds and generates
a unitary that reduces the energy of the system. The energy taken
from the system is stored in a classical field or in a battery and
is referred to as work. In a four-stroke engine the strokes are thermalization
of the hot manifold, unitary evolution, thermalization of the cold
manifold, and another unitary evolution. In the continuous machine
all terminals (baths and work repository) are connected simultaneously:
hot bath, cold bath, and battery/classical field. In this paper we
shall refer to this machine as simultaneous and not continuous for
reasons explained later on. 

Due to the above mentioned non commutativity, different machines operate
in a different manner, and in general their performances differ (even
in cases where they have the same efficiency as in the numerical examples
in \cite{EquivPRX}) . Nonetheless, the thermodynamic equivalence
principle presented in \cite{EquivPRX} states that in the quantum
limit of small action, all machine types are thermodynamically equivalent.
That is, they have the same work per cycle, and the same heat flows
per cycle. This equivalence takes place where the operation of each
stroke is very close to the identity operation. This regime is characterized
by 'engine action' that is small compared to $\hbar$. This does not
mean low power since a small action cycle can be completed in a short
time. Regardless of how close to identity the operation are, the different
machine types exhibit very different dynamics (for example the simultaneous
machine does not have a pure unitary stage). Nevertheless, the equivalence
principle states that all these differences disappear when looking
on the \textit{total heat or total work after an integer number of
cycles}. The details of the equivalence principle will become clear
as we present our results for the non Markovian case.

The paper is organized as follows. Section I describes the engine
and baths setup and introduces the heat exchangers. In Sec. II we
derive the equivalence relation in the non-Markovian regime (short
evolution time or strong couplings). The equivalence of heat machine
types is valid when the ``engine action'' is small compared to $\hbar$.
In Sec. III it is shown how to choose the initial state of the batteries
in the weak action regime so that their entropy will not change in
the charging process. In addition, we find that for large action a
different initial battery state is preferable. In the end of the section
it is shown that for some initial states of the battery the machine
charges the battery with energy while reducing the entropy of the
battery at the same time. Section IV shows that in contrast to Markovian
machines it is possible to construct machines with a higher degree
of equivalence. The emphasis is on the very existence of such machines,
because their usefulness is presently unclear. In Sec. V we conclude.

\section{The setup}

We describe here the minimal model needed for extending the equivalence
principle to short time dynamics beyond the Markovian approximation.
However, the same logic and tools can be applied to more complicated
systems with more levels or more baths. The setup studied in this
paper is shown in Fig. 1. Heat is transferred from the bath to the
engine (black ellipse) via particles of the heat exchangers (circles
on gears). In each engine cycle the gears turn and a fresh particle
enters the interaction zone (gray shaded area) where it stays until
the next cycle. The engine can only interact with the particles in
the interaction zone. The work repository is a stream of particles,
or batteries, (green circles) that are 'charged' with work by the
engine. This interaction of the elements with the engine can be turned
on and off as described by the periodic functions $f_{k}(t)=f_{k}(t+\tau_{cyc})$
where $\tau_{cyc}$ is the machine cycle time. Throughout the paper
we shall use the index 'k' as a 'terminal index' that can take the
values 'h','c' or 'w' that stand for hot, cold and work repository
(battery) respectively. In what follows we elaborate on the different
elements in the scheme. 

\begin{figure}
\includegraphics[width=8.6cm]{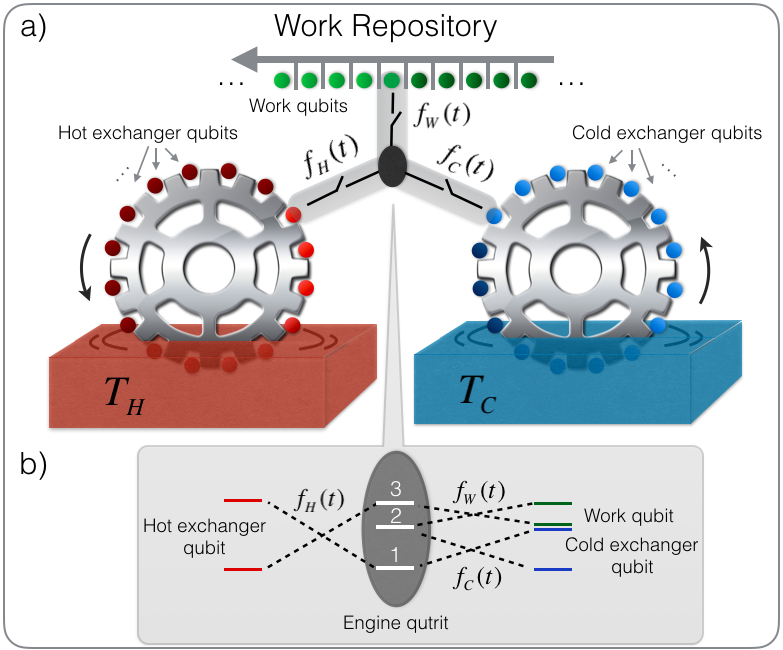}

\caption{a) Heat machine scheme with heat exchangers (gears). Various engine
types can be implemented in this scheme by controlling the coupling
function $f_{c,h,w}(t)$ to the engine (ellipse). In each cycle the
gears turn and the work repository shifts so that new particles enter
the interaction zone (gray shaded area). The heat exchangers enable
the use of Markovian baths while having non Markovian engine dynamics.
This includes strong coupling and/or short time evolution. In this
model the work is stored in many batteries (work qubit in green).
b) The engine level diagram. This machine is based on two-body energy
conserving unitaries. This is in contrast to other machines that employ
three-body interaction.}
\end{figure}

\subsection{The heat exchanger and the baths}

The heat machine equivalence principle \cite{EquivPRX} calls for
a small engine action which implies in the present formalism $\nrm{H_{ek}}\tau\ll\hbar$.
However, in the microscopic derivation of the Lindblad equation a
rotating wave approximation is made. The approximation is valid only
if $\tau$ is large compared to the oscillation time. This implies
that $\nrm{H_{ek}}$ has to be very small in the equivalence regime.
Nevertheless, in principle, small action can be achieved with strong
(or weak) coupling $\nrm{H_{ek}}$ and short evolution time as long
as $\nrm{H_{ek}}\tau\ll\hbar$ holds. In this regime, which is the
subject of this paper, the dynamics is highly non Markovian. A non
Markovian bath dynamics is in general very complicated and strongly
depends on the specific bath realization. Heat machines and the second
law in the presence of strong coupling have been discussed in \cite{EisertMachineBeyondWeak,EspositoNonequilibGreen}.

To overcome the complicated dynamics and \textit{to obtain results
that are universal and not bath-realization dependent, we add heat
exchangers to our setup}. Heat exchangers are abundant in macroscopic
heat machines. In house air conditioning a coolant fluid is used to
pump heat form the interior space to an external cooling unit. Water
in a closed system is used in car engines to transfer heat to the
radiator where air can cool the water. Perhaps the simplest example
is the cooling fins that are used to cool devices like computer chips.
The metal strongly interacts with the chip and conduct the heat to
the fins. Then, the weak coupling with the air cools the fins. The
interaction with the air is weak but due to the large surface it accumulates
to large heat transfer that cools the chip. 

In the quantum regime heat exchangers enable the following simplifications.
Firstly, it separates the system interaction scales from the bath
interaction scales. The system can undergo non Markovian dynamics
with the heat exchangers while the baths can thermalize the heat exchangers
using standard weak coupling Markovian dynamics. Secondly, it enables
to start each engine cycle with a known environment state (the particles
in the interaction zone). Most importantly, this environment state
is in a product state with the system and contains no memory of previous
cycles. Thirdly, it eliminates the dependence on all bath parameters
except the temperature. This means, that from the point of view of
the machine, all different bath realizations are equivalent. This
bath parameters independence holds as long as the bath fully thermalizes
the heat exchanger particles.

In our scheme the coolant fluids consist of $N_{c}$ and $N_{h}$
particles in each heat exchanger (particles around the gears in Fig
1a). The particles in the gear cyclically pass through the bath and
the machine interaction zone (gray shaded area) cyclically with periods
of $N_{c,h}\tau_{cyc}$. Note that the gears in Fig 1a are merely
an illustration of the heat exchanger concept. The heat exchanger
can be realized, for example, by adjacent superconducting qubit or
by moving neutral atoms with light. The particles may interact strongly
and in a non Markovian way with the system. On the other hand, the
particles interact \textit{weakly} with the bath but for a sufficiently
long time so that they fully thermalize when they leave the bath.
After the exchanger particles exit the bath they are in a thermal
state and in a product state with the system (the bath removes all
correlation to the machine). In each cycle of the engine a different
exchanger particle interacts with the system. $N_{c,h}$ are analogous
to the size of the 'cooling fin'. Their number is chosen so that within
the Markovian, weak-coupling limit to the baths, for all practical
purposes, they have sufficient time to fully thermalize. 

Under the assumptions above it does not matter what are the exact
details of the Markovian bath (e.g., its thermalization and correlation
time). It only needs to induce thermal state via weak coupling (to
avoid strong interaction issues). In this regime the thermalization
can be described by the Lindblad equation \cite{breuer}. However,
because of the heat exchanger full thermalization assumption, there
is no need for an explicit solution.

The work repository is basically a heat exchanger without a bath.
It may have a conveyor belt geometry as shown in Fig 1a, or it may
be cyclic like the heat exchanger. The considerations of choosing
the initial state of the batteries particles (work repository) will
be discussed later.

The model can be extended by letting the system interact with more
than one heat exchanger particle at a time, or by not fully thermalizing
the particles. However, it seems like these types of extensions eliminate
the advantages of using heat exchangers to begin with. The simple
setup described above is sufficient to exemplify the equivalence principle
in short time non Markovian dynamics.

\subsection{The engine}

The engine core shown in Fig. 1b, is a three-level system. Level 1
and 3 constitute the hot manifold with an energy gap $E_{h}$, levels
1 and 2 constitute the cold manifold with a gap $E_{c}$, and the
work manifold comprises levels 2 and 3. The more general notion of
manifold separation in quantum heat machines is described in \cite{EquivPRX}.

The hot (cold) manifold can interact only with the hot (cold) heat
exchanger. This interaction can be switched on and off without any
energetic cost as explained in the next section. The same holds for
the work repository. If the engine qutrit interacts only with one
heat exchanger, the hot for example, then the hot manifold of the
system will eventually reach a Gibbs state at temperature $T_{h}$. 

For the engine operating regime we want the thermal strokes to create
population inversion that would be used to excite the batteries to
higher energies. This simple engine structure facilitates the construction
of thermal machines using only two-body interactions rather than three-body
interactions \cite{linden10,k272,malabarba2014clock}.%
{}

\subsection{The coupling of the engine to the heat exchangers and to the work
repository}

In our model, the particles in the heat exchangers are all qubits.
The energy gaps of the qubits in the heat exchangers are equal to
the energy gaps of their corresponding manifold in the engine qutrit.
As explained earlier, in each engine cycle the heat exchangers dials
turn and a new thermal particle is available to interact with the
system. These exchanger particles are not initially correlated to
the engine so the initial state (in each cycle) of the particles in
the engine interaction zone is $\rho_{tot}(t=0)=\rho_{c}\otimes\rho_{h}\otimes\rho_{w}\otimes\rho_{e}$
where $\rho_{e}$ is the engine state and $\rho_{c,h,w}$ are the
bath and work repository particles that are in the interaction zone
of the system. The rest of the particles are not required until the
next cycle of the machine.

The coupling between the engine and the hot bath particles has the
form:
\begin{equation}
H_{int}=\sum_{k=c,h,w}f_{k}(t)H_{ek}\label{eq: Hint}
\end{equation}
where $f_{k}(t)$ are the controllable periodic scalar couplings (switches
in Fig. 1a and dashed lines in Fig. 1b) introduced earlier. $H_{ek}$
are energy conserving Hamiltonian: If $H_{k}$ is the Hamiltonian
of the exchanger particle and $H_{e}$ is the qutrit engine Hamiltonian,
then energy conserving interaction satisfy: $[H_{ek},H_{e}+H_{k}]=0$.
This condition is the standard assumption in thermodynamic resource
theory. It is used to define 'thermal operations', and it ensures
that energy is not exchanged with the controller that generates $H_{ek}$.
The total energy in the exchangers, work repositories and the engine
is not affected by $H_{ek}$. Thus, $H_{ek}$ can only redistribute
the total energy but not change it.

The simplest form of $H_{int}$ is $H_{ek}=a_{k}a_{ek}^{\dagger}+a_{k}^{\dagger}a_{ek}$
where $a_{k}$ is the annihilation operator for the k exchanger particle,
and $a_{ek}$ is the annihilation operator for hot manifold in the
engine. These $H_{ek}$ Hamiltonians generate a partial (or a full)
swap between the $k$ manifold in the machine and the terminal $k$.
This operation is slightly more complicated than the standard partial
swap as will be explained in the battery section.

In the beginning of each cycle the engine starts in a product state
with its immediate environment. This inserts a Markovianity scale
to the model since there is no bath memory from cycle to cycle. Nonetheless,
there are still important non-Markovian aspects in the intra-cycle
dynamics. The full Markovian dynamics is obtained in the weak collision
limit \cite{RUswap,rybar2012simulation,ziman2005description,GennaroQuDit,gennaro2008entanglement},
where \textit{in each thermal stroke} the engine interacts weakly
with \textit{many particles of the heat exchanger}.

In the simultaneous machine all the $f_{k}$ are turned on and off
together in order to couple the machine to different particles in
the heat exchangers. Thus, the couplings are not fixed in time as
in the Markovian continuous machine. While Markovian continuous machines
do not have a cycle time, the simultaneous machines have a cycle time
$\tau_{cyc}$ determined by the rate that particles of the heat exchangers
enter the interaction zone.

\subsection{The work repository}

There are two major thermodynamic tasks: one is to produce work, and
the other is to change the temperature of an object of interest. While
cooling can be done either by investing work (power refrigerator)
or by using only heat baths (absorption refrigerator), engines always
involve the production of work. Often the receiver of the work is
not modeled explicitly. Instead a classical field is used to drive
the system and harvest the work. This is equivalent to a repository
that is big and hardly changes its features due to the action of the
engine. 

When the work repository is modeled explicitly various complications
arise. First, the state of the battery may change significantly (especially
if the battery starts close to its ground state) and therefore affect
the operation of the engine (back action). Second, as entanglement
starts to form between the battery and the system, the reduced state
of the battery gains entropy. The energy exchange can no longer be
considered as pure work. In an ideal battery the energy increases
without any accompanying entropy change. This feature is captured
by the entropy pollution measure: $\Delta S/\Delta\left\langle E\right\rangle $
\cite{Woods2015EffEng,CollectiveArxiv}. In a good battery this number
is very small and can even be negative as will be shown later.

To avoid the back action problem we will use multiple batteries. In
the present scheme it is sufficient to use qubits or qutrits. That
is, instead of raising one weight by a large amount, we raise many
weights just a little. In some cases this is indeed the desired form
of work. For example, an engine whose purpose to prepare many particles
with population inversion that are later used as a gain medium for
a laser. 

As with the heat exchanger, the batteries will be connected to the
engine sequentially, one in each cycle with an interaction term of
the form (\ref{eq: Hint}). The reduced state of a terminal particle
$k$ (may belong to the heat exchanger or to the battery) after the
engine operated on it, will be denoted by $\rho_{k}'=tr_{\neq j}[\rho_{tot}]$.
In general, after the cycle the terminal particle may be strongly
and/or classically correlated to the engine.

The initial state of the battery is a key issue that affects dramatically
the entropy pollution and the quality of charging the battery with
work. Nevertheless, it is not directly related to the issue of heat
machine equivalence so we will discuss the battery initial state only
in Sec. III.

\subsection{Heat and work}

The heat that flows into the cold (hot) bath in one cycle is given
by the change in the energy of the heat exchanger particle  after
one cycle:
\begin{equation}
Q_{c(h)}^{cycle}=tr[(U_{cyc}\rho_{tot}(0)U_{cyc}^{\dagger}-\rho_{tot}(0))H_{c(h)}\otimes\unit{_{else}}]\label{eq: Q cyc}
\end{equation}
where $U_{cyc}$ is the evolution operator generated by $H_{int}$
over one cycle of the machine. Writing this in terms of the state
of the whole system rather than using the reduced state of the bath,
is very useful. To evaluate the total change in the bath energy we
need to know the global transformation of over one cycle $U_{cyc}$.
The internal dynamics which is machine dependent has no impact on
the total heat. All engines that have the same $U_{cyc}$ will have
the same amount of heat per cycle. This is in contrast to the Markovian
Lindblad formalism. There, a symmetric rearrangement theorem had to
be applied to show that the total heat per cycle is the same for different
machines. In the present case, when the one cycle evolution is equivalent
for different types of machines, (\ref{eq: Q cyc}) immediately implies
equivalence of heat and work per cycle. 

As for energy exchange with the work repository $\Delta\left\langle H_{w}\right\rangle $,
we replace $H_{c(h)}$ by $H_{w}$ in (\ref{eq: Q cyc}). In order
to identify it with work it is required that no entropy is generated
in the work repository.

\section{the equivalence of heat machines in the non-markovian regime}

The construction of various heat machine types in the same physical
system was studied in \cite{EquivPRX} and it is based on operator
splitting techniques. In particular the Strang splitting \cite{de1987prodStrang,feit1982Strang,StrangDecompError2000jahnke}
for two non commuting operators $A$ and $B$ is $e^{(A+B)dt}=e^{\frac{1}{2}Adt}e^{Bdt}e^{\frac{1}{2}Adt}+O(dt^{3})$.
Starting with the simultaneous machine operator where all terminals
are connected simultaneously:
\begin{eqnarray}
\tilde{U}_{cyc}^{simul} & = & e^{-i[\mc H_{e}+\underset{k=c,h,w}{\sum}\mc H_{k}+\mc H_{ek}]\tau_{cyc}}\nonumber \\
 & = & U_{0}U_{cyc}^{simul},\\
U_{cyc}^{simul} & = & e^{-i[\mc H_{ec}+\mc H_{eh}+\mc H_{ew}]\tau_{cyc}},
\end{eqnarray}
where $U_{0}=e^{-i(\mc H_{e}+\mc H_{c}+\mc H_{h}+\mc H_{w})\tau_{cyc}}$,
the single-particle (SP) coherence evolution operator can be singled
out from the total evolution operator since by virtue the condition
$[\mc H_{e}+\mc H_{c}+\mc H_{h}+\mc H_{w},H_{int}]=0$. All the population
change is generated by $U_{cyc}^{simul}$ \footnote{In fact, $U_{cyc}^{simul}$ is the evolution operator in the interaction
picture. Energy observables like heat look the same in the interaction
picture ($U_{0}^{\dagger}H_{c,h,w}U_{0}=H_{c,h,w}$). In practice
all states should be evolved with $U_{cyc}^{simul}$ only. The bare
Hamiltonians $H_{k}$ are used only for calculating the energy observables
.}. Thus, the SP coherences associated with interaction-free time evolution
$U_{0}$ do not affect the population dynamics and observables like
energy that are diagonal in the energy basis. The fact that $U_{0}$
commutes with $U_{cyc}^{simul}$ means that that outcome of the operation
does not depend on the time the operation is carried out (time invariance). 

This type of SP coherences should be distinguished from inter-particle
(IP) coherences. Since the energy gaps in the machine and the terminal
are matched, the IP coherences are between degenerate states. For
example the states $\ket{0_{c}1_{e}}$ and $\ket{1_{c}0_{e}}$ are
degenerate, and so are the pairs $\{\ket{0_{h}3_{e}},\ket{1_{h}0_{e}}\}$
and $\{\ket{0_{w}3_{e}},\ket{1_{w}2_{e}}\}$. The crossed lines in
Fig. 1b show the pairs of two-particle degenerate states. These IP
coherences are essential for the dynamics. Their complete suppression
leads to a Zeno effect that halts all the dynamics in the engine.
The IP coherences are generated and modified by the interaction terms
and hence cannot be separated from the rest of the evolution like
the SP coherences. Note that changes in IP coherence translate to
population changes in the subspaces of individual particles.

When starting in a product state where the IP coherences are zero,
the energy transfer (population changes) is of order $dt^{2}$ while
the coherence generation is of order $dt$. This is due to the fact
that unitary transformation converts population to coherences and
coherence to population (see Fig. 8 in \cite{EquivPRX}). In thermodynamic
resources theory phases are often dismissed as non-essential but we
stress that this is true only for the SP coherences.

\subsection*{Evolution operator decompositions}

To study the relations between the simultaneous engine and the two-stoke
engine, we apply the Strang decomposition which yields the following
product form
\begin{eqnarray}
U_{cyc}^{simul} & = & e^{-i\mc H_{ew}\frac{\tau_{cyc}}{2}}e^{-i(\mc H_{ec}+\mc H_{eh})\tau_{cyc}}e^{-i\mc H_{ew}\frac{\tau_{cyc}}{2}}+O[(\frac{s}{\hbar}){}^{3}]\nonumber \\
 & = & \mc U^{II\:stroke}+O[(\frac{s}{\hbar}){}^{3}]\label{eq: Uequiv}
\end{eqnarray}
where $s$ is the 'engine norm action' $s=(\nrm{\mc H_{ec}}_{sp}+\nrm{\mc H_{eh}}_{sp}+\nrm{\mc H_{ew}}_{sp})\tau_{cyc}$
and $\nrm{\cdot}_{sp}$ is the spectral norm of the operator \cite{EquivPRX}.
When this number is small compared to $\hbar$, $U_{cyc}^{II-stroke}\to U_{cyc}^{simul}$.
Note that the first term and the third term in $U_{cyc}^{II-stroke}$
are two parts of the same stroke. The operator splits this way since
the Strang splitting can only create symmetric units cell. A similar
splitting can be done for the four-stroke engine exactly as shown
in \cite{EquivPRX}. One immediate conclusion follows from the equivalence
of the one cycle evolution operators: if different machine types \textit{start
in the same initial condition}, their state will coincide when monitored
stroboscopically at $t_{n}=n\tau_{cyc}$. While at $t_{n}=n\tau_{cyc}$
the states of different machine types will differ by $O[(\frac{s}{\hbar}){}^{3}]$
at the most, at other times they will differ by $O[\frac{s}{\hbar}]$.
This expresses the fact that the machine types are never identical
at all times. They differ in the strongest order possible $O[\frac{s}{\hbar}]$,
unless complete cycles are considered. Since the one cycle evolution
operators are equivalent, it follows from (\ref{eq: Q cyc}) that
\textit{for the same initial engine state}:

\begin{equation}
Q_{h(c)}^{simul}\cong Q_{h(c)}^{II\:stroke}\cong Q_{h(c)}^{IV\:stroke}\label{eq: Q equiv}
\end{equation}

where $Q^{simul}$ refers to the heat transferred in time of $\tau_{cyc}$
in the particle machine. The $\cong$ stands for equality up to correction
$\nrm{H_{c(h)}}O[(\frac{s}{\hbar}){}^{4}]$. Note that the cubic term
does not appear in (\ref{eq: Q equiv}). Due to lack of initial coherence
the $O[(\frac{s}{\hbar}){}^{3}]$ correction contribute only to the
IP coherence generation but not to population changes. Hence, the
population changes differ only in order $O[(\frac{s}{\hbar}){}^{4}]$.
In transients the system energy changes from cycle to cycle so in
general it is not correct to use energy conservation to deduce from
heat equality on work equality. Nevertheless, work equality follows
from (\ref{eq: Uequiv}) and (\ref{eq: Q cyc}) when using $H_{w}$
instead of $H_{c(h)}$.

This establishes the equivalence of heat (and work) even very far
away from steady state operation or thermal equilibrium \textit{provided
all engines start with the same state}. This behavior is very similar
to the Markovian equivalence principle \cite{EquivPRX}, but there
is one major difference. Since each cycle starts in a product state
the leading order in heat and work is $O[(\frac{s}{\hbar}){}^{2}]$
and not $O[\frac{s}{\hbar}]$ as in the Markovian case. The linear
term in the work originates from the SP coherence generated by the
classical driving field. Without this coherence the power scales as
$(Q_{h}^{cyc}+Q_{c}^{cyc})/\tau_{cyc}\propto\tau_{cyc}$. Thus, as
shown in Fig. 2a, for small action the power grows linearly with the
cycle time. On the other hand, as explained earlier, the correction
to the power is only of order $O[(\frac{s}{\hbar}){}^{4}]/\tau_{cyc}=O[(\frac{s}{\hbar}){}^{3}]$
since there is no cubic correction to the work.

Let us consider now the \textit{steady state operation}. Despite (\ref{eq: Q equiv}),
it is not immediate that the heat will be the same for different machine
types in steady state. In (\ref{eq: Q equiv}) the initial density
matrix is the same for all machine types. However, different types
may have slightly different initial states, which may affect the total
heat. To study equivalence in steady state operation we first need
to define what steady state means when the bath and batteries are
included in the analysis. The whole system is in a continuous transient:
the hot bath gets colder, the cold bath gets hotter, and the batteries
are charged. Nonetheless, the reduced state of the engine relaxes
to a limit cycle $\bar{\rho}_{e}(t)=\bar{\rho}_{e}(t+\tau)$ or explicitly
$\bar{\rho}_{e}=tr_{\neq e}[U_{cyc}(\rho_{c}\otimes\rho_{h}\otimes\rho_{w}\otimes\bar{\rho}_{e})U_{cyc}]$.
To see the relation between steady states of different machines we
choose the steady state of one machine, for example $\bar{\rho}_{e}^{simul}$,
and apply the two-stroke evolution operator to it:

\begin{eqnarray}
\bar{\rho}_{e}' & = & tr_{\neq e}[U_{cyc}^{II\:stroke}(\rho_{c}\otimes\rho_{h}\otimes\rho_{w}\otimes\bar{\rho}_{e}^{simul})U_{cyc}^{II\:stroke}]\nonumber \\
 & = & \bar{\rho}_{e}^{simul}+O[(\frac{s}{\hbar}){}^{4}]\label{eq: steady state quartic}
\end{eqnarray}
The cubic order is absent because if there are no initial IP coherences,
then the cubic term only generates IP coherences. Hence, the reduced
state of the system is not modified by the cubic correction of the
two-stroke evolution operator. From (\ref{eq: steady state quartic})
we conclude that the steady states are equal for both machines up
to quartic corrections in the engine action. From (\ref{eq: Q equiv})
it follows that heat and work in steady state are also equal in all
machine types, up to quartic correction. Figure 2a shows the power
in steady state for the three main types of machines as well as for
a higher order six-stroke machine that will be discussed in the last
section. Let the power of the machine (work per cycle divided by cycle
time) be denoted by $P$. In Fig 2b, we plot the normalized power
$P/P^{simul}$ where it is easier to see that the correction in the
power of one machine \textit{with respect to the other} is quadratic.
This graph shows that the equivalence of non Markovian machines is
actually similar to that of Markovian machines. The difference is
that the reference simultaneous power is constant (in action) in the
Markovian case and linearly growing (small action) in the present
case. Fig. 2b shows that the equivalence is a phenomenon that takes
place in a regime and not only at the (ill defined) point $\tau_{cyc}=0$.
The same holds for the Markovian case.

At this point we wish to discuss the quantumness of the equivalence
principle in the current setup. In \cite{EquivPRX} it was suggested
to use dephasing in the energy basis to see if the machine is stochastic
or quantum. If a dephasing stroke is carried out before the unitary
stroke and the result is not affected, then the machine operates as
a stochastic machine. In the four-stroke engine and in the two-stroke
engine described in (\ref{eq: Uequiv}) the battery is accessed twice
during the cycle. The first interaction with the battery creates some
IP coherence between the battery and the engine. As a result, the
next interaction with the battery (the second work stroke) starts
with nonzero IP coherence. Thus, adding dephasing after the first
work stroke will affect the power gained in the next work stroke.
This is shown by the red and blue dashed curves in Fig 2a. The power
of the simultaneous engine is zero if we continuously dephase the
system (Zeno effect). We conclude that although there is no coherence
that carries over from one cycle to the next, as in the Markovian
case, coherence is still needed for the equivalence principle to hold.
This time the coherence is an IP coherence between degenerate states.

Until this point we ignored the nature of energy transferred to the
battery, i.e. if it is heat or work. If it is pure work the device
is an engine, whilst if it is heat the device functions as an absorption
machine (only heat bath terminals). However, the equivalence principle
is indifferent to this distinction. If the action is small, two-stroke,
four-stroke, and simultaneous machines will perform the same. In the
next section, however, we study the conditions under which the entropy
of the batteries is not increased and the device performs as a proper
engine.

\begin{figure}
\includegraphics[width=8.6cm]{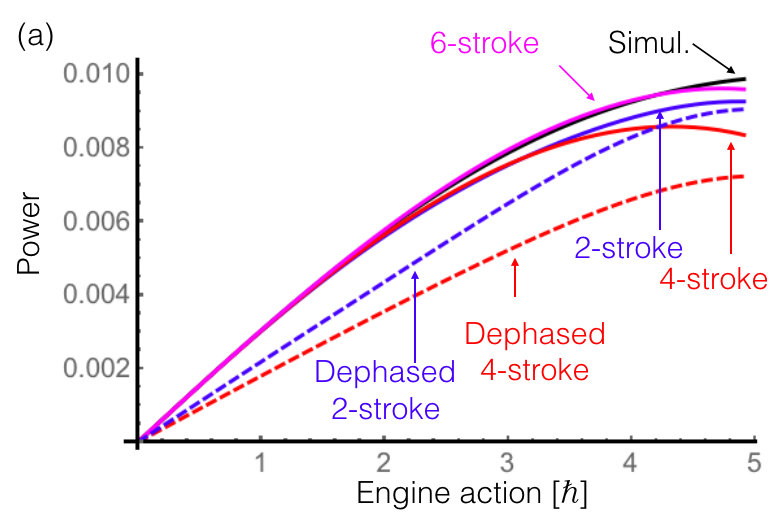}

\includegraphics[width=8.6cm]{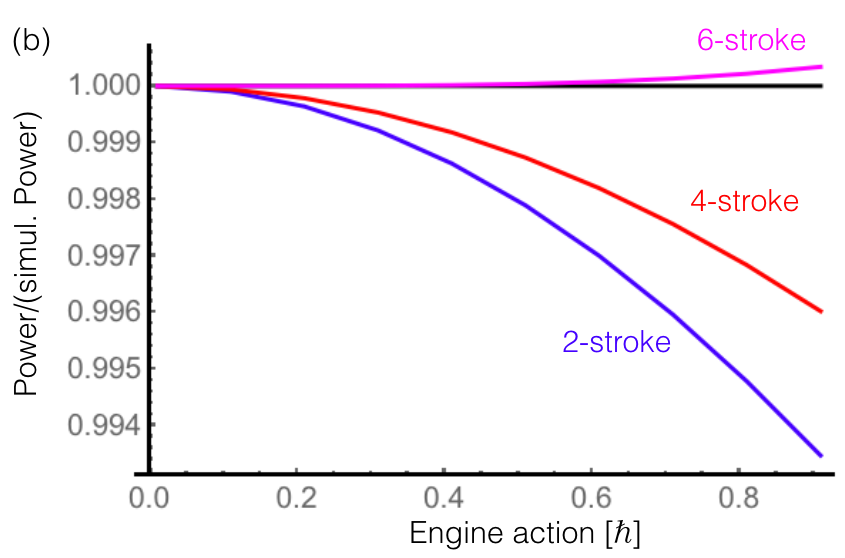}

\caption{(a) In the non Markovian regime the main machine types: four-stroke,
two-stroke and simultaneous machine, have the same power when the
engine action is small compared to $\hbar$. In contrast to the Markovian
case here the power is not constant but grows linearly for small action.
The action is increased by increasing the time duration of each stroke.
The red and blue dashed curves show how the 4-stroke and 2-stroke
engines are modified when a dephasing stroke is included. This demonstrates
that the thermodynamic equivalence is a quantum coherence effect.
(b) The equivalence become more visible when plotting the relative
power of each machine with respect to the simultaneous machine. The
6-stroke machine, based on the Yoshida decomposition, is unique to
the non-Markovian case and has a wider range of equivalence.}

\end{figure}

\section{Work extraction}

\subsection{The initial state of the battery in strong and weak coupling}

So far, we have not explicitly addressed the question of work extraction
and whether the energy transferred to the batteries is actually pure
work or heat. For engines, the goal is to make the entropy pollution
$\Delta S_{w}/\Delta\left\langle H_{w}\right\rangle $ as small a
possible. Figure 3 shows the well knows expression for the entropy
of a qubit as a function of the excited state population $p_{w}$.
The von Neumann entropy and the Shannon entropy of single particles
are identical since there are no SP coherences. The energy of the
battery is proportional to the excited state population, so the $x$
axis also indicates the energy of the battery. If the battery starts
with a well defined energy state, that is the ground state $p_{w}=0$,
then a small increase in the energy will result in a large entropy
generation in the battery. In fact, for small changes this is the
worst starting point (the origin in Fig. 3). However, if we choose
to start with a very hot battery at $p_{w}=1/2$, ($T_{w}\to\infty$)
the entropy increase will be very small if $\Delta p_{w}$ is small.
Thus, by using many batteries in a fully mixed state where each is
only slightly changed ($\Delta p_{w}\ll1$) it is possible to reach
the $\Delta S_{w}/\Delta\left\langle H_{w}\right\rangle \to0$ limit.
This is in accordance with the claims that $T_{w}\to\infty$ limit
of an absorption refrigerator, is analogous to a power refrigerator
\cite{levy212}. The price for this choice of the initial state of
the battery is that the number of batteries diverges as $\Delta S_{w}/\Delta\left\langle H_{w}\right\rangle \to0$. 

In the semi-classical field approximation, the field generates a unitary
operation that does not change the entropy of the system. This is
often addressed as pure work as there is no entropy change in the
system. However when modeling the classical field explicitly, one
finds that the source of the classical field actually gains some entropy.
To counter this effect the battery has to be prepared in a special
state \cite{aaberg2014catalytic,malabarba2014clock} or to apply a
feedback scheme \cite{AmikamFlywheel}. Here we suggest to do the
exact opposite and to apply an interaction that will generate a unitary
transformation on the battery but will generate some entropy in the
engine. Consider point B in Fig. 3, a full swap to Point C will increase
the energy but will leave the entropy fixed. In general this will
\textit{increase the entropy of the machine}. Let the initial state
of the engine be $\rho_{e}=diag\{a,b,c\}$ and the initial state of
the work qubit be $\rho_{w}=diag\{1-p_{w},p_{w}\}$. After a \textit{full
swap} interaction we get:
\begin{eqnarray}
\begin{pmatrix}a\\
 & b\\
 &  & c
\end{pmatrix}_{e},\begin{pmatrix}1-p_{w}\\
 & p_{w}
\end{pmatrix}_{w}\to\nonumber \\
\begin{pmatrix}a\\
 & (1-a)(1-p_{w})\\
 &  & (1-a)p_{w}
\end{pmatrix}_{e},\nonumber \\
\begin{pmatrix}b+a(1-p_{w}) & 0\\
0 & c+ap_{w}
\end{pmatrix}_{w}.\label{eq: full swap}
\end{eqnarray}
If $a=0$ a regular full swap takes place between levels 2 and 3 of
the engine and level 1 and 2 of the battery. If $a=1$ there is no
population in level 2\&3 so nothing happens and levels 1\&2 of the
battery remain unchanged. This rule follows from the condition that
guaranties energy conservation $\rho_{e}'-\rho_{e}=-(\rho_{w}'-\rho_{w})$.
Any population change in one particle must be compensated by an opposite
change in the other particle (the energy levels are equal in our model).
Now we demand that this transformation of the battery will generate
a full swap, that is $\rho_{w}'=diag\{p_{w},1-p_{w}\}$. This leads
to the condition $c+ap_{w}=1-p_{w}$ or
\begin{eqnarray}
p_{w} & = & \frac{1-c}{1+a}\label{eq: pw cond}
\end{eqnarray}

Note that $p_{w}$ defines a temperature through the Gibbs factor:
$p_{w}/(1-p_{w})=exp(-E_{w}/T_{w})$. After the full swap the temperature
of levels 2 \& 3 of the engine, is now $T_{w}$. It is simple to show
from the positivity of the quantum mutual information that the entropy
of the engine has increased. This entropy increase is associated with
the formation of correlation (for the full swap it is strictly classical).
When the total population on the subspace of interest on both sides
is not equal (e.g. $a\neq0$ in the example above), classical correlation
forms. If the engine is measured, the marginal distribution of the
battery changes. Another way to see the presence of correlation is
the following. The unitary conserves the total entropy. However the
entropy of the reduced state of the battery does not change while
the reduced entropy of the engine does change. This implies that the
mutual information is larger than zero. This classical correlation
formation can be avoided by replacing the qubit batteries with qutrit
batteries whose initial state is $\rho_{w}=diag\{a,c,b\}$ (note the
flip of $b$ and $c$). In this case the full swap operation will
not generate any correlation between the engine and the battery. 

The full swap is a strong coupling operation. Here, strong coupling
was used to make a more efficient battery charging mechanism compared
to the $T_{w}\to\infty$ alternative in the weak coupling limit.

\begin{figure}
\includegraphics[width=8.6cm]{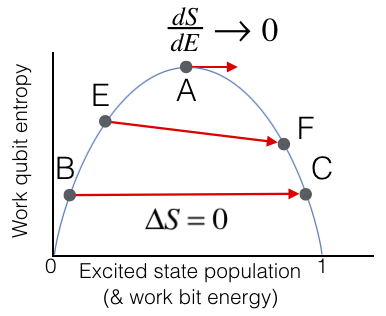}

\caption{For infinitesimal changes (weak coupling) it is preferable to start
with a battery qubit that is close to the fully mixed state (point
$A$) where $dS/dE=0$. For larger changes it is preferable to generate
a permutation that conserves the entropy and creates population inversion
($B\to C$ line). While $\Delta S=0$ for the battery is analogous
to classical field work repository, in two-level batteries it is possible
to charge the batteries and reduce their entropy simultaneously ($E\to F$). }
\end{figure}

\subsection{Beyond the semi-classical limit of the driving field}

When the work repository is described approximated by a classical
field no entropy accounting is carried out for the work repository.
However, for an explicit battery the possible changes in the entropy
of the battery have to be studied. In this subsection it is shown
that these changes can actually be useful. As illustrated in the $E\to F$
trajectory in Fig. 3, it is possible to increase the energy while
reducing the entropy. Strictly speaking, this operation mode does
not exactly correspond to an engine, since the energy change in the
battery is associated with an entropy change as well. Nevertheless,
this change in entropy is a welcomed one, as entropy reduction is
hard to achieve, and often requires some additional resources. In
Fig. 4 we show an explicit example for an initial engine state with
populations $\{a,b,c\}=\{0.056,0.074,0.4\}$ as a function of the
initial exited state population of the battery $p_{w}$. In the shaded
area the energy of the battery is increased while its entropy is reduced.
The left boundary of the shaded regime is given by the $dS=0$ condition
(\ref{eq: pw cond}). The right boundary is given by the condition
$\Delta E=0$. Using (\ref{eq: full swap}) $\Delta E=0$ leads to
the right boundary condition $p_{w}=c/1-a=c/(b+c)$. This condition
means that population ratio in the engine qutrit and in the battery
is the same. Hence, nothing happens when the swap is carried out.
This zero change in population also leads to $\Delta S=0$ at this
point.

\begin{figure}
\includegraphics[width=8.6cm]{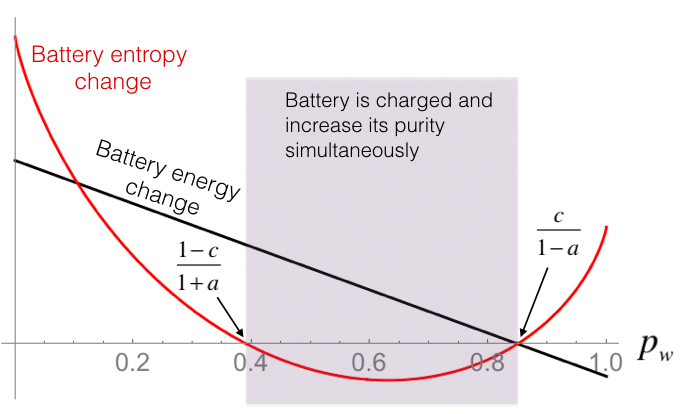}\caption{For a given engine the changes in energy (black) and entropy (red)
of the battery are plotted as a function of the initial excited state
probability $p_{w}$ of the battery. a,b and c are the population
of level 1,2 and 3 of the engine just before the interaction with
the battery starts. The shaded area shows the regime where the battery
is not only charged but also purified. This can only be done by strong
interaction with the engine and the battery.}

\end{figure}

\section{Higher order splittings}

The regime of equivalence studied above and in \cite{EquivPRX} is
determined by the use of the Strang decomposition for the evolution
operator. Although higher order decompositions do exist, they involve
coefficients with alternating signs \cite{Blanes2005necessNegCoef}.
In the Markovian case this is not physical since a bath that generates
evolution of the form $\exp(-\mc Lt)$ is not physical (where $\exp(+\mc Lt)$
is the standard Lindblad evolution). In the present paper, instead
of non unitary evolution of the reduced state of the engine, we consider
the global evolution operator of all the components. The global evolution
is unitary and its generators, the interaction Hamiltonian, are all
Hermitian. Hence, there is no problem to have for example $\exp(+iH_{ec}t)$
instead of $\exp(-iH_{ec}t)$. It simply means an interaction term
with opposite sign. This facilitates the use of higher order decompositions
in order to make machines with more strokes and a larger regime or
equivalence. 

In \cite{Yoshida1990construction} Yoshida introduced a very elegant
method to construct higher order decompositions. Let $U_{s^{2}}(t)=U^{simul}(t)+O[(\frac{s}{\hbar}){}^{4}]$
stand for an evolution operator that has a correction of order $s^{3}$
with respect to $U^{simul}(t)$. It can be, for example, a four-stroke
or two-stroke engine. As shown in \cite{Yoshida1990construction},
a fourth order evolution operator $U_{s^{4}}(t)$ can be constructed
from $U_{s^{2}}(t)$ in the following way:
\begin{eqnarray}
U_{s^{4}}(t) & = & U_{s^{4}}(x_{1}t)U_{s^{4}}(x_{0}t)U_{s^{4}}(x_{1}t)\label{eq: Yoshida 4}\\
\{x_{0},x_{1}\} & = & \{\frac{-2^{1/3}}{1-2^{1/3}}\:,\:\frac{1}{1-2^{1/3}}\}\nonumber 
\end{eqnarray}
where $U_{s^{4}}(t)=U^{simul}(t)+O[(\frac{s}{\hbar}){}^{5}]$. By
applying the same arguments as before, when the cycle starts with
fresh uncorrelated bath and battery particles the correction to the
work and heat are $O[(\frac{s}{\hbar}){}^{6}]$. The Yoshida method
is powerful since it can be repeated, with different $x_{0},x_{1}$
coefficients, to gain operators that are even closer to the simultaneous
machine. Physically, (\ref{eq: Yoshida 4}) can be interpreted as
a regular $U_{s^{2}}(t)$ machine where the stroke durations alternate
every cycle. Figure 2b shows the ratio of the power of various engines
with respect to the simultaneous engine. While in the Strang four-stroke
and two-stroke machine the \textit{power} deviation from the simultaneous
machine is second order in the action, the power of the Yoshida engine
of order four deviates from the simultaneous machine only in the fourth
order in the action.

Two-stroke and four-stroke engines naturally emerge from practical
considerations. Two-stroke engines emerge when it is easier to thermalize
simultaneously the hot and cold manifolds. Four-stroke engines emerge
when it is easier to thermalize one manifold at a time. In contrast,
the Yoshida decomposition (\ref{eq: Yoshida 4}) does not split the
simultaneous engine into more basic or simpler operations compared
to the two-stroke and four-stroke machines. Thus, the practical motivation
for actually constructing Yoshida-like higher order machines is not
obvious at all at this point. Nevertheless, our main point in this
context is that higher order machines are forbidden in Markovian dynamics
and are allowed in the non Markovian machines studied here.

\section{Conclusion}

It has been demonstrated that the principle of thermodynamics equivalence
of heat machine types is valid beyond Markovianity. We find higher
order equivalence relations that do not exist in the Markovian regime.
In addition it was shown that the strong coupling limit enables to
deliver finite work to the battery without increasing its entropy.
It also enables to charge and reduce the entropy of the battery at
the same time. In our setup we introduced heat exchangers to mediate
between the machine and the baths. Heat exchangers significantly simplify
the analysis, but they also have a significant practical value. They
remove the strong dependence on the finer properties of the baths,
and allow more flexible machine operating regimes while still using
a simple Markovian bath.
\begin{acknowledgments}
This work was supported by the Israeli science foundation. Part of
this work was supported by the COST Action MP1209 'Thermodynamics
in the quantum regime'. RU cordially thanks the support the Kenneth
Lindsay trust fund.
\end{acknowledgments}

\bibliographystyle{apsrev4-1}
\bibliography{/Users/raam_uzdin/Dropbox/RaamCite}

\end{document}